\documentclass[twocolumn,secnumarabic,amssymb,showpacs,superscriptaddress, nobibnotes, aps, prb]{revtex4-1}

\usepackage{graphicx}   
\usepackage{dcolumn}   
\usepackage{bm}           
\usepackage{amssymb}   

\begin{document}

\title{Semiconductor, topological semimetal, indirect semimetal, and topological Dirac semimetal phases of Ge$_{1-x}$Sn$_{x}$ alloys}%

\author{H.-S. Lan}%
\email[Corresponding author : ]{huangsianglan@gmail.com}
\affiliation{Graduate Institute of Electronics Engineering and Department of Electrical Engineering, National Taiwan University, Taipei, Taiwan}
\author{S. T. Chang}%
\affiliation{Department of Electrical Engineering, National Chung Hsing University, Taichung, Taiwan}
\author{C. W. Liu}%
\email[Corresponding author : ]{chee@cc.ee.ntu.tw}
\affiliation{Graduate Institute of Electronics Engineering and Department of Electrical Engineering, National Taiwan University, Taipei, Taiwan}
\affiliation{Graduate Institute of Photonics and Optoelectronics, National Taiwan University, Taipei, Taiwan}
\affiliation{National Nano Device Labs, Hsinchu, Taiwan}
\date{February 28, 2017}%

\begin{abstract}
Electronic structures of Ge$_{1-x}$Sn$_{x}$ alloys (0 $\leq$ $x$ $\leq$ 1) are theoretically studied by nonlocal empirical pseudopotential method. For relaxed Ge$_{1-x}$Sn$_{x}$, a topological semimetal is found for $\textit x$ $>$ 41$\%$ with gapless and band inversion at ${\Gamma}$ point, while there is an indirect-direct bandgap transition at $x$ = 8.5$\%$. For strained Ge$_{1-x}$Sn$_{x}$ on a Ge substrate, semimetals with a negative indirect bandgap appear for $x$ $>$ 43$\%$, and the strained Ge$_{1-x}$Sn$_{x}$ on Ge is always an indirect bandgap semiconductor for $x$ $<$ 43$\%$. With appropriate biaxial compressive strains, a topological Dirac semimetal is found with band inversion at ${\Gamma}$ and one pair of Dirac cones along the [001] direction.
\end{abstract}
\maketitle

Gray tin ($\alpha$-Sn) is a topological semimetal\cite{d1,d2,d3} (also referred to topological zero-gap semiconductor) due to its inverted $\Gamma_{7}^{-}$ and $\Gamma_{8}^{+}$ band states at zone center in the reciprocal space, satisfying the non-zero Z$_2$ topological invariant in the $k_{\text{z}}$ = 0 plane.\cite{d3} Moreover, a strong topological insulator in strained $\alpha$-Sn was proposed with an opening gap between the split ${\Gamma}_{8}^{+}$ band states ($\Gamma_{8}^{+}$ and $\Gamma_{8}^{+*}$) at zone center due to the unchanged parity eigenvalues under biaxial strain.\cite{d3} The Dirac semimetal\cite{d4}  with one pair of Dirac cones in the $\Gamma$-Z direction in compressive strained $\alpha$-Sn was then reported.\cite{d5,d6} Relaxed Ge has the same parity eigenvalues as relaxed $\alpha$-Sn for the four occupied bands at eight time-reversal invariant momenta (1$\Gamma$, 3X, and 4L) except the degenerate $\Gamma_{8}^{+}$ band states at zone center.\cite{d7} However, if band inversion occurred ($\Gamma_{7}^{-}$ energy level lower than $\Gamma_{8}^{+}$), the odd parity of $\Gamma_{7}^{-}$ would lead to the non-zero Z$_2$ invariant. Note that these parity eigenvalues are used to identify the Z$_2$ invariant, where 0 and 1 for relaxed Ge and $\alpha$-Sn in the $k_{\text{z}}$ = 0 plane, respectively.\cite{d3} Recently, diamond structure GeSn alloys have been attractive for light-emitting\cite{d8,d9,d10} and electronics\cite{d11,d12} applications owing to the potential direct bandgap of GeSn alloys and small transport effective masses, respectively. The indirect-direct bandgap transition of relaxed GeSn (r-GeSn) alloys reportedly occurred at Sn content around 7${-}$10\%.\cite{d8,d13,d14,d15,d16} A zero gap behavior was also reported for r-GeSn at the Sn content larger than ${\sim}$40\% based on band structure calculations.\cite{d17,d18} However, the occurrence of the band inversion in metallic GeSn alloys, implying a non-zero Z$_2$ invariant in GeSn alloys, has not been discussed yet. The nonlocal empirical pseudopotential method (EPM) has been widely used for calculating the electronic band structures of SiGe\cite{d19,d20,d21} and GeSn\cite{d11,d15,d22} alloy systems using virtual crystal approximation (VCA). The pseudocharge density, $\rho$$_{\text{pseu}}$, calculated by the electronic wave function was used for determining the bonding characteristics of Ge\cite{d23} and $\alpha$-Sn\cite{d24}. Noted that the $\Gamma_{7}^{-}$ and $\Gamma_{8}^{+}$ band states at zone center are {\em s}-like (antibonding {\em s} orbitals) and {\em p}-like (bonding {\em p} orbitals), respectively, for both Ge and $\alpha$-Sn.\cite{d25} The calculated bandgaps and band offsets in s-GeSn/r-GeSn using our EPM, where Sn content $\leq$ 0.3, have been reported.\cite{d26} In this work, the phase transition from a semiconductor to a topological semimetal in r-Ge$_{1-x}$Sn$_{x}$ alloys (0 $\leq$ $x$ $\leq$ 1) is investigated using EPM. This transition is determined from the corresponding wave functions of $\Gamma_{7}^{-}$ and $\Gamma_{8}^{+}$ band states at zone center. For strained Ge$_{1-x}$Sn$_{x}$ (s-Ge$_{1-x}$Sn$_{x}$) alloys (0 $\leq$ $x$ $\leq$ 1), three phases (semiconductor, indirect semimetal, and topological Dirac semimetal) are found depending on Sn content and compressive strain level.

In EPM, the one-electron pseudo-Hamiltonian derived from Ref.~\onlinecite{d27} has four terms of the kinetic energy, local pseudopotential form factors ({\em V}$_{\text{loc}}$({\em q})), nonlocal correction terms ({\em V}$_{\text{nloc}}$($\vec{G}$, $\vec{G}$$^{\prime}$)), and spin-orbit interactions ({\em V}$_{\text{so}}$($\vec{G}$, $\vec{G}$$^{\prime}$)). The {\em V}$_{\text{loc}}$({\em q}) versus reciprocal lattice vectors ({\em q} = ${|}${\em G} ${-}$ {\em G}$^{\prime}$${|}$) are presented by the expression\cite{d19,d28} 
\begin{equation}
V_{\text{loc}}^{i}(q) = \frac{1}{\Omega^{i}} \frac{b_{\text{1}}^{i}(q^{2}{-}b_{\text{2}}^{i})}{\exp{[b_{\text{3}}^{i}(q^{2}{-}b_{\text{4}}^{i})]}+1}\biggl(\frac{1}{2}\tanh{\Bigl(\frac{b_{\text{5}}^{i}{-}q^{2}}{b_{\text{6}}^{i}}\Bigr){+\frac{1}{2}}}\biggr), \label{pauli}
\end{equation}
where ${\Omega^{i}}$ is the atomic volume and {\em i} denotes the Ge or Sn element. The parameters of  {\em b}$_{\text{1}}^{i}$, {\em b}$_{\text{2}}^{i}$, {\em b}$_{\text{3}}^{i}$, and {\em b}$_{\text{4}}^{i}$ are obtained by solving roots of a system of nonlinear equations with the values of {\em V}$_{\text{loc}}^{i}$({\em q}) at {\em q}$^{2}$ = $\{$3, 4, 8, 11$\}$ ${\times}$ (2${\pi}$/{\em a}$_{\text{0}}^{i}$)$^{2}$. The lattice constants ({\em a}$_{\text{0}}$) at 0 K of Ge (5.652 {\AA}) and Sn (6.482 {\AA}) are calculated using the value at RT\cite{d29} and the corresponding thermal expansion coefficients\cite{d30}. EPM parameters of {\em V}$_{\text{loc}}$({\em q}), spin-orbit interactions (${\zeta}$ and ${\mu}$), and a fast cut-off “tanh” part\cite{d28} ({\em b}$_{\text{5}}^{i}$ and {\em b}$_{\text{6}}^{i}$) of Ge and $\alpha$-Sn (Table I) are adopted from Refs.~\onlinecite{d20,d22,d31,d32} with less than 6\% adjustment to reach good agreement with the experimental bandgaps of Ge\cite{d29,d33} and $\alpha$-Sn\cite{d34,d35,d36} at low temperature. The parameters of nonlocal correction terms are obtained from Ref.~\onlinecite{d27} and the number of element plane wave basis set $\{$$\vec{G}$$\}$ is 339. Details of the three terms ({\em V}$_{\text{loc}}$, {\em V}$_{\text{nloc}}$, and {\em V}$_{\text{so}}$) were reported comprehensively by theoretical works.\cite{d15,d19,d20,d27,d31,d32,d37} Here, we describe the approaches to take into account both strain and alloy effects in these three terms. The terms of {\em V}$_{\text{loc}}$ and {\em V}$_{\text{nloc}}$, and the parameter ${\lambda}$ in the {\em V}$_{\text{so}}$ of Ge-Sn alloy systems are obtained by VCA using the following formulas
\begin{equation}
V^{\text{Ge}_{\text{1-{\em x}}}\text{Sn}_{\text{{\em x}}}}(q) = (1{-}x)\frac{\Omega^{\text{Ge}}}{\Omega_{\text{s}}^{\text{Ge}_{\text{1-{\em x}}}\text{Sn}_{\text{{\em x}}}}}V^{\text{Ge}}(q){+}x\frac{\Omega^{\text{Sn}}}{\Omega_{\text{s}}^{\text{Ge}_{\text{1-{\em x}}}\text{Sn}_{\text{\em x}}}}V^{\text{Sn}}(q), \label{pauli}
\end{equation}
{\text{both for}} {\em V}$_{\text{loc}}$ {\text{and}} {\em V}$_{\text{nloc}}$.
\begin{equation}
{\lambda}^{\text{Ge}_{\text{1-{\em x}}}\text{Sn}_{\text{{\em x}}}}(K, K^{\prime}) = (1{-}x){\lambda}^{\text{Ge}}(K, K^{\prime})+x{\lambda}^{\text{Sn}}(K, K^{\prime}). \label{pauli}
\end{equation}
The $\{$$\vec{G}$$\}$ and the normalizing strained atomic volume ${\Omega_{\text{s}}^{i}}$ generated from the lattice vectors are considered in the Hamiltonian matrix\cite{d20} with the strain and alloy effects. The linear interpolation of elastic constants ({\em C}$_{\text{11}}$, {\em C}$_{\text{12}}$, and {\em C}$_{\text{44}}$)\cite{d38} and a bowing of 0.047 {\AA}\cite{d39} for the lattice constant of GeSn alloys are used. Note that the coherent potential approximation (CPA) in agreement with the VCA results in homogeneous GeSn alloys (substitutional $\alpha$-Sn in Ge) was reported.\cite{d40,d41} CPA was used to consider the inhomogeneous GeSn alloys with $\beta$-Sn defects, that may not exist at zero Kelvin discussed in this work according to the formula in Ref.~\onlinecite{d41}.
 
The band structure of r-Ge$_{0.65}$Sn$_{0.35}$ [Fig. 1(a)], a typical direct-gap semiconductor, owns the conduction band edge at zone center ($\Gamma_{7}^{-}$ state) and the degenerate valence band edges ($\Gamma_{8}^{+}$ states for heavy hole and light hole bands) with bandgap of ${\sim}$70 meV. However, the band structure of r-Ge$_{0.55}$Sn$_{0.45}$ [Fig. 1 (b)] shows a gapless topological semimetal behavior with the degenerate $\Gamma_{8}^{+}$ states above the $\Gamma_{7}^{-}$ state. The corresponding constant $\rho_{\text{pseu}}$ contours around the two atoms in the unit cell of the band states of {\em s}-like $\Gamma_{7}^{-}$ and {\em p}-like $\Gamma_{8}^{+}$ show the band inversion of r-Ge$_{0.55}$Sn$_{0.45}$ as compared to r-Ge$_{0.65}$Sn$_{0.35}$ [Fig. 1 (c) and (d)]. The band inversion leads to a non-zero Z$_2$ invariant as referred to the topological behavior. The same parity eigenvalues of r-Ge$_{0.55}$Sn$_{0.45}$ as r-Sn for the four occupied bands at eight time-reversal invariant momenta are confirmed using our EPM. Without the spin-orbit coupling (SOC), the parity inversion in r-Ge$_{0.55}$Sn$_{0.45}$ is disappeared.

\begin{table}
\caption{\label{tab:fonts}Pseudopotential parameters used for Ge and Sn. (Ref.~\onlinecite{d31}).}
\begin{ruledtabular}
\begin{tabular}{lcc} 
Parameter & Ge & Sn \\
\colrule
\text{\em {V}}$_{\text{loc}}$($\sqrt{3}$)(Ry) & $-$0.2351\footnote{Reference~\onlinecite{d32}.} & $-$0.191 \\
\text{\em {V}}$_{\text{loc}}$($\sqrt{4}$)(Ry) & $-$0.1572 & $-$0.152 \\
\text{\em {V}}$_{\text{loc}}$($\sqrt{8}$)(Ry) & 0.0186\footnotemark[1] & $-$0.008 \\
\text{\em {V}}$_{\text{loc}}$($\sqrt{11}$)(Ry) & 0.055\footnotemark[1] & 0.04 \\
${\zeta}$(${\AA}$$^{-1}$) & 5.34 & 4.75 \\
${\mu}$ (10$^{-4}$ Ry) & 9.4\footnote{Reference~\onlinecite{d20}.} & 22.5 \\
\text{\em {b}}$_{\text{5}}$ (atomic units) & 4.5\footnotemark[1] & 3.9\footnote{Reference~\onlinecite{d22}.} \\
\text{\em {b}}$_{\text{6}}$ (atomic units) & 0.3\footnotemark[1] & 0.3\footnotemark[3] \\
\end{tabular}
\end{ruledtabular}
\end{table}

\begin{figure}
\includegraphics{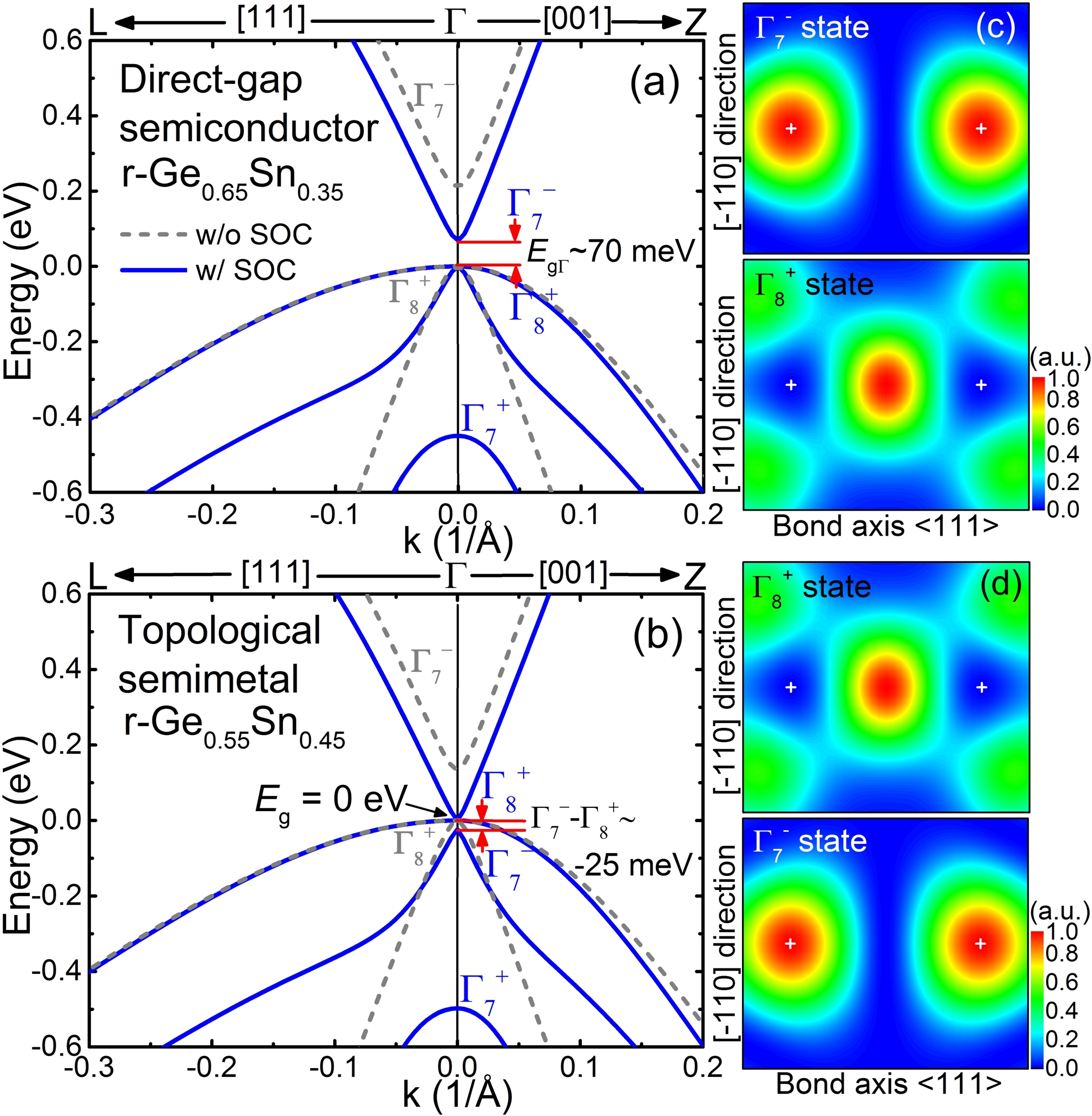} 
\caption{\label{fig:epsart} The calculated band structures with and without spin-orbit coupling (SOC) around zone center of (a) a direct-gap semiconductor (r-Ge$_{0.65}$Sn$_{0.35}$) and (b) a gapless topological semimetal (r-Ge$_{0.55}$Sn$_{0.45}$). The constant pseudocharge density ($\rho$$_{\text{pseu}}$) contours of the two atoms (indicated by ${+}$) in the unit cell of (c) r-Ge$_{0.65}$Sn$_{0.35}$ and (d) r-Ge$_{0.55}$Sn$_{0.45}$ to identify {\em s}-like $\Gamma_{7}^{-}$ and {\em p}-like $\Gamma_{8}^{+}$.}
\end{figure}

\begin{figure}
\includegraphics{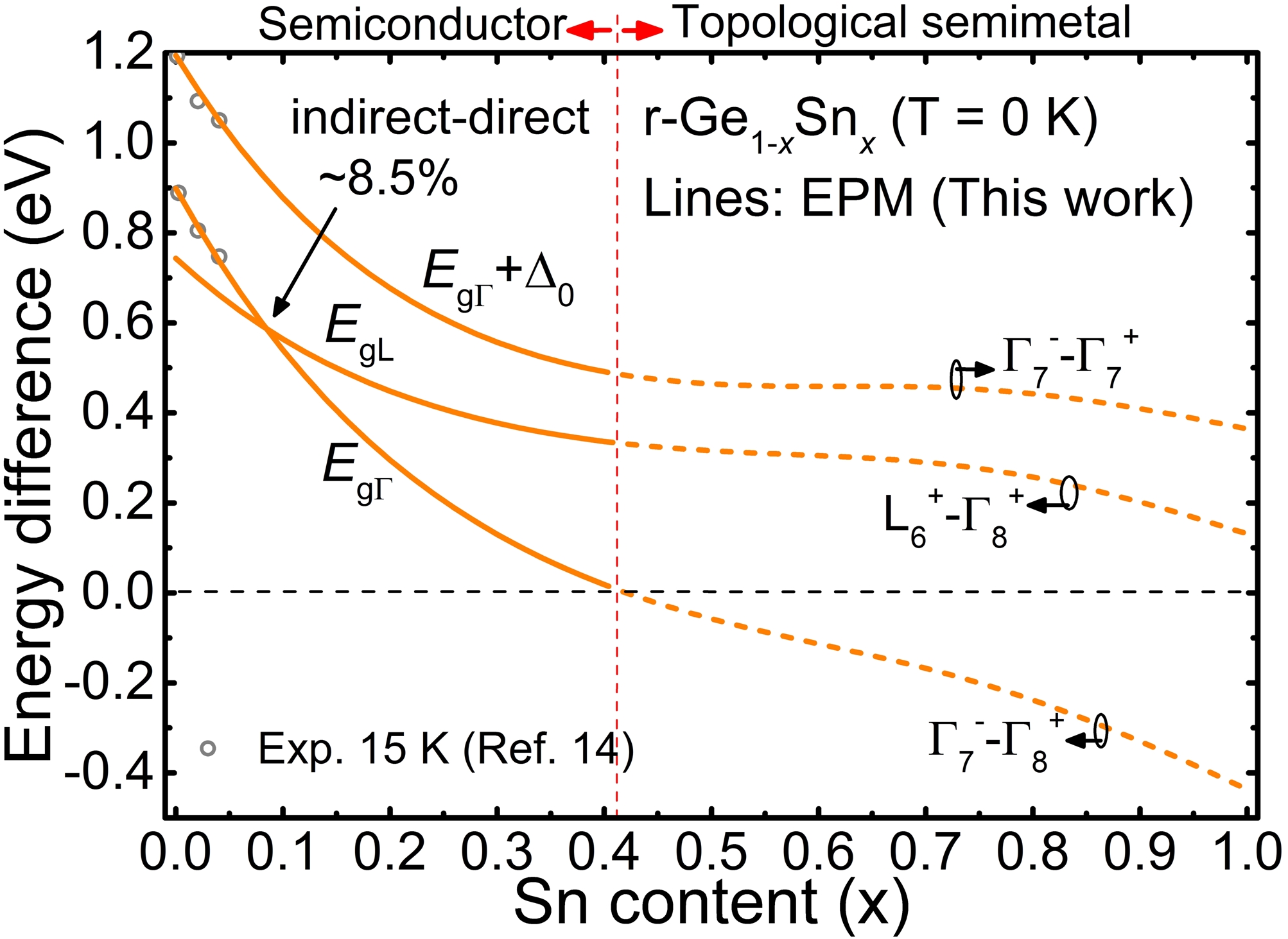} 
\caption{\label{fig:epsart} The calculated energy differences of $\Gamma_{7}^{-}$-$\Gamma_{8}^{+}$, L$_{\text{6}}^{+}$-$\Gamma_{8}^{+}$, and $\Gamma_{7}^{-}$-$\Gamma_{7}^{+}$ as a function of Sn content as compared with the reported experimental data (Ref.~\onlinecite{d14}) (Permission granted from CCC). The solid lines indicate the semiconductor bandgaps including {\em E}$_{\text{gL}}$, {\em E}$_{\text{g}{\Gamma}}$, and {\em E}$_{\text{g}{\Gamma}}$ + ${\Delta}$$_{\text{0}}$, and the dashed lines ({\em x} ${>}$ 41\%) indicate the energy differences of the topological semimetal.}
\end{figure}

The calculated indirect bandgap ({\em E}$_{\text{gL}}$ = L$_{\text{6}}^{+}$-$\Gamma_{8}^{+}$), direct bandgap ({\em E}$_{\text{g}{\Gamma}}$ = $\Gamma_{7}^{-}$-$\Gamma_{8}^{+}$), and {\em E}$_{\text{g}{\Gamma}}$ + spin-orbit splitting (${\Delta}$$_{\text{0}}$) ({\em E}$_{\text{g}{\Gamma}}$+${\Delta}$$_{\text{0}}$ = $\Gamma_{7}^{-}$-$\Gamma_{7}^{+}$) as a function of Sn content for r-Ge$_{1-x}$Sn$_{x}$ are shown in Fig. 2. Our calculations agree well with the reported experimental data at low Sn content ({\em E}$_{\text{g}{\Gamma}}$ and {\em E}$_{\text{g}{\Gamma}}$+${\Delta}$$_{\text{0}}$) near 0 K\cite{d14,d42}. The calculated bandgaps of $\alpha$-Sn ({\em E}$_{\text{gL}}$ = 0.13 eV, {\em E}$_{\text{g}{\Gamma}}$ = $-$0.43 eV, and ${\Delta}$$_{\text{0}}$ = 0.8 eV) are also consistent with reported values ({\em E}$_{\text{gL}}$ = 0.12 eV,\cite{d34,d36} {\em E}$_{\text{g}{\Gamma}}$ = $-$0.42 eV,\cite{d35} and ${\Delta}$$_{\text{0}}$ = 0.8 eV\cite{d35}). There is no experimental {\em E}$_{\text{gL}}$ data of r-Ge$_{1-x}$Sn$_{x}$ near 0 K reported in literature. {\em E}$_{\text{g}{\Gamma}}$ decreases faster than {\em E}$_{\text{gL}}$ with increasing Sn content and this results in an indirect-direct bandgap transition around {\em x} = 8.5\% for r-Ge$_{1-x}$Sn$_{x}$.\cite{d26} For x $>$ 41\%, the degenerate $\Gamma_{8}^{+}$) forms a gapless topological semimetal ({\em E}$_{\text{g}}$ = 0 eV) with the band inversion that the {\em s}-like $\Gamma_{7}^{-}$ falls below the two {\em p}-like $\Gamma_{8}^{+}$ states in energy, i.e., $\Gamma_{7}^{-}$-$\Gamma_{8}^{+}$ ${\sim}$$-$25 meV of r-Ge$_{0.55}$Sn$_{0.45}$ [Fig. 1 (b)]. The energy differences $\Gamma_{7}^{-}$-$\Gamma_{8}^{+}$, L$_{\text{6}}^{+}$-$\Gamma_{8}^{+}$, and $\Gamma_{7}^{-}$-$\Gamma_{7}^{+}$ in gapless r-Ge$_{1-x}$Sn$_{x}$ alloys are also shown in Fig. 2 for comparison.

\begin{figure}
\includegraphics{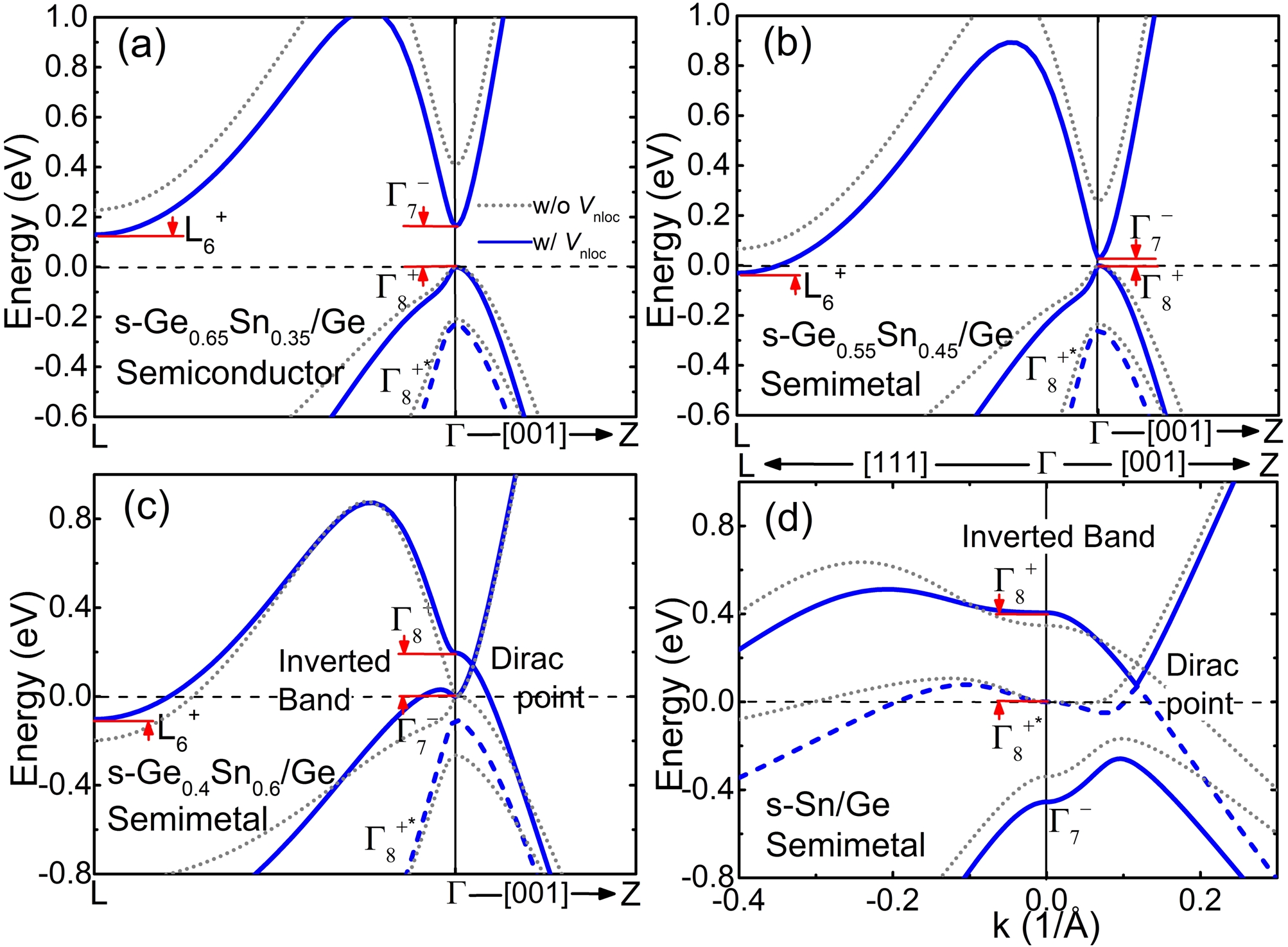} 
\caption{\label{fig:epsart} Electronic band structures of s-Ge$_{1-x}$Sn$_{x}$ on Ge with different Sn content with and without nonlocal potential ($V_{\text{nloc}}$). (a) An indirect-gap semiconductor (s-Ge$_{0.65}$Sn$_{0.35}$ on Ge). (b) An indirect semimetal (s-Ge$_{0.55}$Sn$_{0.45}$ on Ge). (c) An indirect semimetal with the inverted band at zone center and a Dirac point along the [001] direction (s-Ge$_{0.4}$Sn$_{0.6}$). (d) An indirect semimetal with inverted band at zone center and a Dirac point along the [001] direction (s-Sn on Ge). Note that the opening gap at zone center is $\Gamma_{8}^{+}$-$\Gamma_{7}^{-}$ and $\Gamma_{8}^{+}$-$\Gamma_{8}^{+*}$ for (c) and (d), respectively.}
\end{figure}

For s-Ge$_{1-x}$Sn$_{x}$ layer pseudomorphiclly grown on a Ge (001) substrate, the phase transition in the band structure from semiconductor to indirect semimetal with the increase of Sn content is shown in Fig. 3. Note that we assume that a metastable fully strained thin layer s-Ge$_{1-x}$Sn$_{x}$ could be grown on Ge even though a high Sn content of s-Ge$_{1-x}$Sn$_{x}$ on Ge (x $>$ 46\%) is still under investigation.\cite{d43,d44} The s-Ge$_{0.65}$Sn$_{0.35}$ on Ge [Fig. 3 (a)] has a typical indirect bandgap with conduction band edges at L$_{\text{6}}^{+}$ states and the valence band edge at $\Gamma_{8}^{+}$ state (the heavy hole band). For s-Ge$_{0.55}$Sn$_{0.45}$ on Ge [Fig. 3 (b)], the L$_{\text{6}}^{+}$ states fall below the $\Gamma_{8}^{+}$ state, resulting in an indirect semimetal with a negative indirect bandgap (L$_{\text{6}}^{+}$-$\Gamma_{8}^{+}$ ${\sim}$$-$30 meV). As Sn content reaches to 60\% [Fig. 3 (c)], the band inversion of $\Gamma_{7}^{-}$ and $\Gamma_{8}^{+}$ states occurs at zone center with an opening gap ($\Gamma_{8}^{+}$-$\Gamma_{7}^{-}$) at $\Gamma$ point and a Dirac point along the [001] direction, but the L$_{\text{6}}^{+}$ states are still the conduction band minimum. In this case, s-Ge$_{0.4}$Sn$_{0.6}$ on Ge is referred to an indirect semimetal with a negative indirect bandgap (not a topological Dirac semimetal) owing to the uncertainly occupied $\Gamma_{7}^{-}$ state with respect to the unknown Fermi energy.\cite{d6}  For s-Sn on Ge in Fig. 3 (d), the large compressive strain (${\sim}$$-$12.8\%) moves the $\Gamma_{8}^{+*}$ state upwards beyond the $\Gamma_{7}^{-}$ state. However, conduction band edges remain at L$_{\text{6}}^{+}$ states. Moreover, the Dirac points are along [001] direction, not along [100] or [010] direction on the compressively strained plane, which is consistent with the previous report.\cite{d45} Without the energy dependence $V_{\text{nloc}}$ term for the core states, the symmetries allow the occurrences of band inversion and Dirac point even though loss of accuracy in energy. The coexistence of band inversion and Dirac point in s-Ge$_{0.4}$Sn$_{0.6}$ on Ge without $V_{\text{nloc}}$ is shown in Fig. 3 (c).

Fig. 4 shows detailed phase transitions in s-Ge$_{1-x}$Sn$_{x}$ on Ge as a function of Sn content. The calculated energies, {\em E}$_{\text{gL}}$, $\Gamma_{8}^{+*}$-$\Gamma_{8}^{+}$, and $\Gamma_{7}^{+}$-$\Gamma_{8}^{+}$, of s-Ge$_{1-x}$Sn$_{x}$ on Ge are consistent with reported experimental data\cite{d46,d47} at low Sn content (not shown). For 0 $\leq$ {\em x} $\leq$ 30\%, {\em E}$_{\text{g}{\Gamma}}$ decreases faster than {\em E}$_{\text{gL}}$. As a result, the energy difference {\em E}$_{\text{g}{\Gamma}}$-{\em E}$_{\text{gL}}$ decreases with increasing Sn content. However, no crossover point is found because the increasing biaxial compressive strain with increasing Sn content moves the $\Gamma_{7}^{-}$ state upwards as compared to the L$_{\text{6}}^{+}$ states, and thus the difference ({\em E}$_{\text{g}{\Gamma}}$$-${\em E}$_{\text{gL}}$) increases again for {\em x} ${>}$ ${\sim}$30\%. An indirect semimetal with a negative indirect bandgap, L$_{\text{6}}^{+}$-$\Gamma_{8}^{+}$, occurs for {\em x} ${>}$ 43\%. The band inversion at zone center is found for {\em x} ${>}$ 47\%, and the opening gap at zone center changes from $\Gamma_{8}^{+}$-$\Gamma_{7}^{-}$ to $\Gamma_{8}^{+}$-$\Gamma_{8}^{+*}$ at {\em x} ${\sim}$ 68\% due to the upward movement of $\Gamma_{8}^{+*}$ energy beyond the $\Gamma_{7}^{-}$ state with increasing biaxial compressive strain.

\begin{figure}
\includegraphics{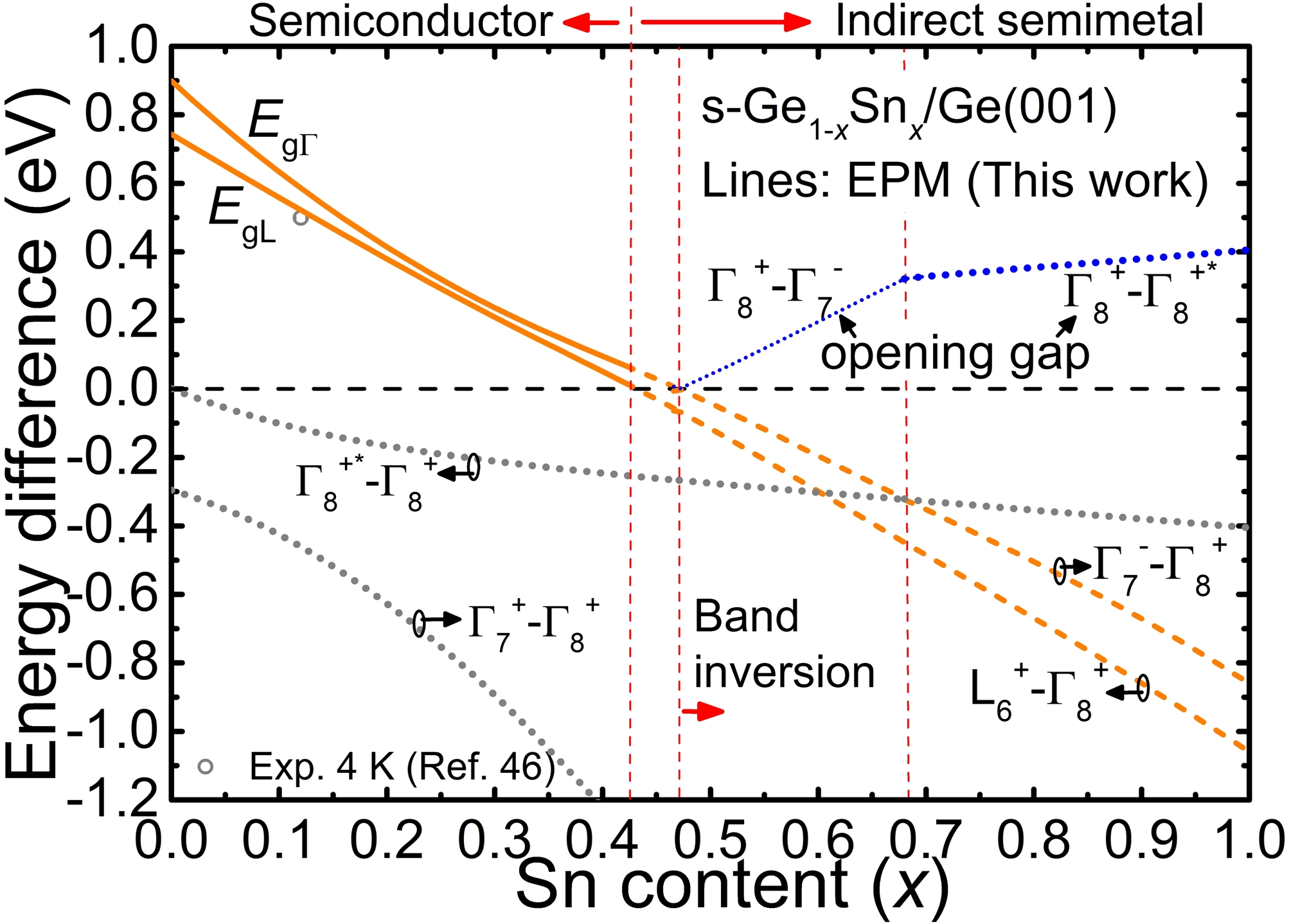} 
\caption{\label{fig:epsart} The calcuated energy differences of $\Gamma_{7}^{-}$-$\Gamma_{8}^{+}$, L$_{\text{6}}^{+}$-$\Gamma_{8}^{+}$, $\Gamma_{8}^{+*}$-$\Gamma_{8}^{+}$, and $\Gamma_{7}^{+}$-$\Gamma_{8}^{+}$ of s-Ge$_{1-x}$Sn$_{x}$ on Ge as a function of Sn content as compared with the reported experimental data (Ref.~\onlinecite{d46}) (Permission granted from CCC). The semiconductor to indirect semimetal transition is found at {\em x} ${>}$ 43\% and the band inversion at zone center occurs at  {\em x} ${>}$ 47\%. The opening gap at zone center changes from $\Gamma_{8}^{+}$-$\Gamma_{7}^{-}$ to $\Gamma_{8}^{+}$-$\Gamma_{8}^{+*}$ at {\em x} ${\sim}$ 68\%.}
\end{figure}

\begin{figure*}
\includegraphics{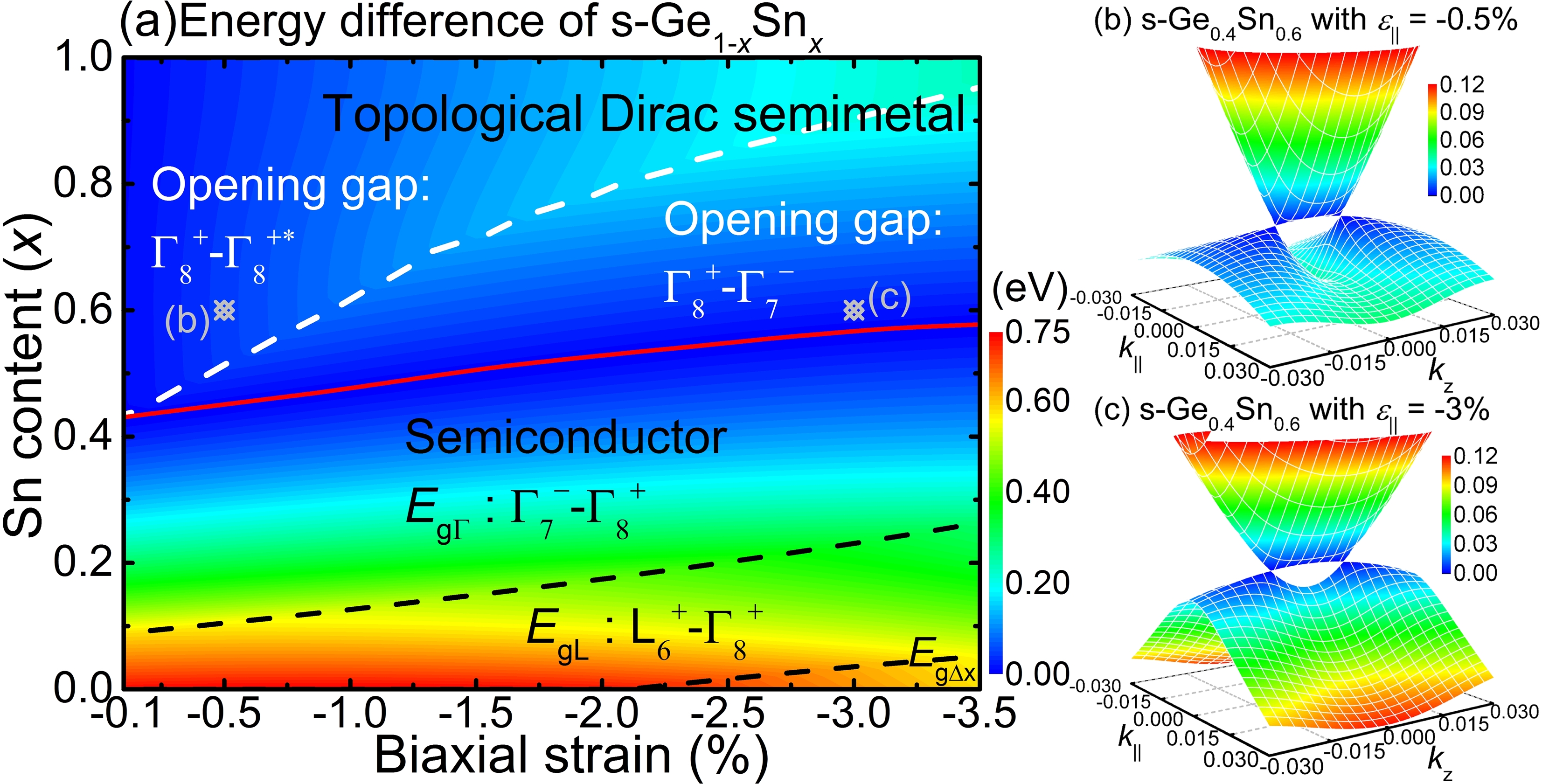} 
\caption{\label{fig:epsart} (a) The phase diagram as a function of Sn content (0 $\leq$ $x$ $\leq$ 1) and biaxial compressive strain (0.1\% $\leq$ $|{\varepsilon}_{||}|$ $\leq$ 3.5\%). The red line indicates the phase transition between direct-gap semiconductor and topological Dirac semimetal. The black dash lines distinguish the three regions for the fundamental bandgap ({\em E}$_{\text{g}}$$_{\Delta}$$_{\text{x}}$, {\em E}$_{\text{gL}}$, and {\em E}$_{\text{g}{\Gamma}}$) of semiconductor. The white dash line distinguishes the two regions for the opening gap ($\Gamma_{8}^{+}$-$\Gamma_{8}^{+*}$ and $\Gamma_{8}^{+}$-$\Gamma_{7}^{-}$) at zone center of topological Dirac semimetal. The band structure on the {\em k}$_{||}$${-}${\em k}$_{\text{z}}$ plane has one pair of three-dimensional Dirac cones located along the {\em k}$_{\text{z}}$ direction for s-Ge$_{0.4}$Sn$_{0.6}$ with (b) the biaxial compressive strain (${\varepsilon}$$_{||}$) of $-$0.5\% and (c) the biaxial compressive strain (${\varepsilon}$$_{||}$) of $-$3\%. The opening gap in (b) is $\Gamma_{8}^{+}$-$\Gamma_{8}^{+*}$, while the opening gap in (c) is $\Gamma_{8}^{+}$-$\Gamma_{7}^{-}$.}
\end{figure*}

In order to form a topological Dirac semimetal in s-Ge$_{1-x}$Sn$_{x}$, the biaxial compressive strain should be smaller than that of s-Ge$_{1-x}$Sn$_{x}$ on Ge to make the energy of L$_{\text{6}}^{+}$ states in Fig. 3 (b) and (c) beyond the $\Gamma_{8}^{+}$ state. In this case, the Fermi energy lies in the middle of Dirac points to ensure the occupied $\Gamma_{7}^{-}$ state.\cite{d3,d6} In the phase diagram defined by Sn content (0 $\leq$ $x$ $\leq$ 1) and biaxial compressive strain (0.1\% $\leq$ $|{\varepsilon}_{||}|$ $\leq$ 3.5\%) [Fig. 5 (a)], the semiconductor/topological Dirac semimetal transition for s-Ge$_{1-x}$Sn$_{x}$ is found in the Sn content range of 41$-$60\% (the red line in Fig. 5 (a)). For semiconductor phase, the fundamental bandgap has three distinct regions for {\em E}$_{\text{g}}$$_{\Delta}$$_{\text{x}}$, {\em E}$_{\text{gL}}$, and {\em E}$_{\text{g}{\Gamma}}$. Note that the conduction band minima at ${\Delta}$ points are split into the 4 fold (2${\Delta}$$_{\text{x}}$ and 2${\Delta}$$_{\text{y}}$) and 2 fold (2${\Delta}$$_{\text{z}}$) valley degeneracies under biaxial compressive strain and the 4 fold has lower energy than the 2 fold. The band structures on the {\em k}$_{||}$${-}${\em k}$_{\text{z}}$ plane of s-Ge$_{0.4}$Sn$_{0.6}$ with the biaxial compressive strain (${\varepsilon}_{||}$) of $-$0.5\% and $-$3\% show one pair of three-dimensional Dirac cones along the {\em k}$_{\text{z}}$ direction [Fig. 5 (b) and (c)] and the band inversion at zone center. Note that the {\em k}$_{||}$ direction refers to the {\em k}$_{\text{x}}$ [100] or {\em k}$_{\text{y}}$ [010] axis perpendicular to the {\em k}$_{\text{z}}$ [001] direction. The non-zero Z$_2$ topological invariant in the {\em k}$_{\text{z}}$ = 0 plane of s-Ge$_{0.4}$Sn$_{0.6}$ with ${\varepsilon}_{||}$ = $-$0.5\% and $-$3\% are also confirmed by the parity eigenvalues of the four occupied bands using our EPM. This is classified as a topological Dirac semimetal.\cite{d4} In addition, the effective Hamiltonian for the Dirac fermion\cite{d4} is used to obtain the velocity along ${k}_{||}$ direction of 8.35$\times10^{6}$ and 1.45$\times10^{7}$ cm$^{2}/$s for s-Ge$_{0.4}$Sn$_{0.6}$ with ${\varepsilon}_{||}$ = $-$0.5\% and $-$3\%, respectively. The opening gap at zone center of topological Dirac semimetal changes from $\Gamma_{8}^{+}$-$\Gamma_{8}^{+*}$ to $\Gamma_{8}^{+}$-$\Gamma_{7}^{-}$ (the white dash line in Fig. 5 (a)) with increasing biaxial compressive strain due to the $\Gamma_{7}^{-}$ energy beyond the $\Gamma_{8}^{+*}$ state under high strain level. 
	
	Semiconductors with a direct or indirect bandgap, indirect semimetals with a negative indirect bandgap, topological semimetals, and topological Dirac semimetals are found in Ge$_{1-x}$Sn$_{x}$ alloy systems by band structure calculations using nonlocal EPM. The Sn content and strain level determine the phase of Ge$_{1-x}$Sn$_{x}$. The existence of diverse phases in Ge$_{1-x}$Sn$_{x}$ alloys has encouraged the exploration of possible phenomena such as chirality, and applications of GeSn alloys.

\begin{acknowledgments}
This work was supported by Ministry of Science and Technology, Taiwan, R.O.C under Grant Nos. 105-2622-8-002-001-, 105-2911-I-009-301, and 103-2221-E-002-232-MY3. The support of high-performance computing facilities by the Computer and Information Networking Center, National Taiwan University, is also highly appreciated.
\end{acknowledgments}


\begin{thebibliography}{99}
\bibitem{d1} Hsin Lin, Tanmoy Das, Yung Jui Wang, L. A. Wray, S.-Y. Xu, M. Z. Hasan, and A. Bansil, Phys. Rev. B, {\bf 87}, 121202(R) (2013).
\bibitem{d2} Julien Vidal, Xiuwen Zhang, Vladan Stevanovi\'{c}, Jun-Wei Luo, and Alex Zunger, Phys. Rev. B, {\bf 86}, 075316 (2012).
\bibitem{d3} Liang Fu and C. L. Kane, Phys. Rev. B, {\bf 76}, 045302 (2007).
\bibitem{d4} Bohm-Jung Yang and Naoto Nagaosa, Nature Communications, {\bf 5}, 4898 (2014).
\bibitem{d5} Victor A. Rogalev, Tomáš Rauch, Markus R. Scholz, Felix Reis, Lenart Dudy, Andrzej Fleszar, Marius-Adrian Husanu, Vladimir N. Strocov, Jürgen Henk, Ingrid Mertig, Jörg Schäfer, and Ralph Claessen, e-print, arXiv:1701.03421 (2017).
\bibitem{d6} Huaqing Huang and Feng Liu, e-print, arXiv:1612.00987 (2016).
\bibitem{d7} M. L. Cohen and J. Chelikowsky, Electronic Structure and Optical Properties of Semiconductors, 2nd edn., Springer Ser. Solid-State Sci. Vol. 75 (Springer, Berlin, Heidelberg, 1989).
\bibitem{d8} S. Wirths, R. Geiger, N. von den Driesch, G. Mussler, T. Stoica, S.Mantl, Z. Ikonic, M. Luysberg, S. Chiussi, J. M. Hartmann, H. Sigg, J. Faist, D. Buca, and D. Grützmacher, Nature Photon., {\bf 9}, 88 (2015).
\bibitem{d9} H. S. Mączko, R. Kudrawiec, and M. Gladysiewicz, Sci. Rep., {\bf 6}, 34082 (2016).
\bibitem{d10} Chung-Yi Lin, Chih-Hsiung Huang, Shih-Hsien Huang, Chih-Chiang Chang, C. W. Liu, Yi-Chiau Huang, Hua Chung, and Chorng-Ping Chang, Appl. Phys. Lett., {\bf 109}, 091103 (2016).
\bibitem{d11} H.-S. Lan and C. W. Liu, Appl. Phys. Lett., {\bf 104}, 192101 (2014).
\bibitem{d12} Yee-Chia Yeo, Xiao Gong, Mark J. H. van Dal, Georgios Vellianitis, and Matthias Passlack., Tech. Dig.-Int. Electron Devices Meet. 2015, 2.4.1.
\bibitem{d13} J. D. Gallagher, C. L. Senaratne, J. Kouvetakis, and J. Menéndez, Appl. Phys. Lett., {\bf 105}, 142102 (2014).
\bibitem{d14} Vijay R. DCosta, Candi S. Cook, A. G. Birdwell, Chris L. Littler, Michael Canonico, Stefan Zollner, John Kouvetakis, and Jos\'{e} Men\'{e}ndez, Phys. Rev. B, {\bf 73}, 125207 (2006).
\bibitem{d15} Kain Lu Low, Yue Yang, Genquan Han, Weijun Fan, and Yee-Chia Yeo, J. Appl. Phys., {\bf 112}, 103715 (2012).
\bibitem{d16} F. L. Freitas, J. Furthm\"{u}ller, F. Bechstedt, M. Marques, and L. K. Teles, Appl. Phys. Lett., {\bf 108}, 092101 (2016).
\bibitem{d17} David W. Jenkins and John D. Dow, Phys. Rev. B, {\bf 36}, 7994 (1987).
\bibitem{d18} Ming-Hsien Lee, Po-Liang Liu, Yung-An Hong, Yen-Ting Chou, Jia-Yang Hong, and Yu-Jin Siao, J. Appl. Phys., {\bf 113}, 063517 (2013).
\bibitem{d19} M. V. Fischetti and S. E. Laux, J. Appl. Phys., {\bf 80}, 2235 (1996).
\bibitem{d20} Martin M. Rieger and P. Vogl, Phys. Rev. B, {\bf 48}, 14276 (1993).
\bibitem{d21} P. Vogl, M. M. Rieger, J. A. Majewski, and G. Abstreiter, Phys. Scr., {\bf T49}, 476 (1993).
\bibitem{d22} Suyog Gupta, Blanka Magyari-K\"{o}pe, Yoshio Nishi, and Krishna C. Saraswat, J. Appl. Phys., {\bf 113}, 073707 (2013).
\bibitem{d23} John P. Walter and Marvin L. Cohen, Phys. Rev. B, {\bf 4}, 1877 (1971).
\bibitem{d24} H. Aourag, B. Soudini, B. Khelifa, and A. Belaidi, Phys. Stat. Sol. (b), {\bf 161}, 685-695 (1990).
\bibitem{d25} P. Y. Yu and M. Cardona, {\sl Fundamentals of Semiconductors}, 3rd ed. (Springer, Berlin, 2001).
\bibitem{d26} H.-S. Lan and C. W. Liu, J. Phys. D: Appl. Phys., {\bf 50}, 13LT02 (2017).
\bibitem{d27}	James R. Chelikowsky and Marvin L. Cohen, Phys. Rev. B, {\bf 14}, 556 (1976).
\bibitem{d28} P. Friedel, M. S. Hybertsen, and M. Schluter, Phys. Rev. B, {\bf 39}, 7974 (1989).
\bibitem{d29} S. Adachi, {\sl Properties of Group-IV, III–V and II–VI Semiconductors} (JohnWiley ${\&}$ Sons, 2005).
\bibitem{d30} R. Roucka, Y.-Y. Fang, J. Kouvetakis, A. V. G. Chizmeshya, and J. Men\'{e}ndez, Phys. Rev. B, {\bf 81}, 245214 (2010).
\bibitem{d31} W. Potz and P. Vogl, Phys. Rev. B, {\bf 24}, 2025 (1981).
\bibitem{d32} Jiseok Kim and Massimo V. Fischetti, J. Appl. Phys., {\bf 108}, 013710 (2010).
\bibitem{d33}	Solomon Zwerdling, Benjamin Lax, Laura M. Roth, and Kenneth J. Button, Phys. Rev., {\bf 114}, 80 (1959).
\bibitem{d34} R. F. C. Farrow, D. S. Robertson, G. M. Williams, A. G. Cullis, G. R. Jones, I. M. Young, and P. N. J. Dennis, J. Cryst. Growth., {\bf 54}, 507 (1981).
\bibitem{d35} T. Brudevoll, D. S. Citrin, M. Cardona, and N. E. Christensen, Phys. Rev. B, {\bf 48}, 8629 (1993).
\bibitem{d36}	R. J. Wagner and A. W. Ewald, J. Phys. Chem. Solids, {\bf 32}, 697 (1971).
\bibitem{d37}	Paul Harrison, {\sl Quantum Well, Wires and Dots}, 2nd Edition (WILEY, 2005), p.373-379.
\bibitem{d38}	Guo-En Chang, Shu-Wei Chang, and Shun Lien Chuang, IEEE J. Quantum Electron., {\bf 46}(12), 1813 (2010).
\bibitem{d39}	R. Beeler, R. Roucka, A. V. G. Chizmeshya, J. Kouvetakis, and J. Menendez, Phys. Rev. B, {\bf 84}, 035204 (2011).
\bibitem{d40}	J. D. Querales-Flores, C. I. Ventura, J. D. Fuhr, and R. A. Barrio, J. Appl. Phys., {\bf 120}, 105705 (2016).
\bibitem{d41}	C. I. Ventura, J. D. Fuhr, and R. A. Barrio, Phys. Rev. B, {\bf 79}, 155202 (2009).
\bibitem{d42}	H. P\'{e}rez Ladr\'{o}n de Guevara, A. G. Rodríguez, H. Navarro-Contreras, and M. A. Vidal, Appl. Phys. Lett., {\bf 91}, 161909 (2007).
\bibitem{d43} Akihiro Suzuki, Osamu Nakatsuka, Shigehisa Shibayama, Mitsuo Sakashita, Wakana Takeuchi, Masashi Kurosawa, and Shigeaki Zaima, Appl. Phys. Lett., {\bf 107}, 212103 (2015).
\bibitem{d44} Akihiro Suzuki, Osamu Nakatsuka, Shigehisa Shibayama, Mitsuo Sakashita, Wakana Takeuchi, Masashi Kurosawa, and Shigeaki Zaima, Jpn. J. Appl. Phys., {\bf 55}, 04EB12 (2016).
\bibitem{d45} A. Barfuss, L. Dudy, M. R. Scholz, H. Roth, P. H\"{o}pfner, C. Blumenstein, G. Landolt, J. H. Dil, N. C. Plumb, M. Radovic, A. Bostwick, E. Rotenberg, A. Fleszar, G. Bihlmayer, D. Wortmann, G. Li, W. Hanke, R. Claessen, and J. Sch\"{a}fer, Phys. Rev. Lett., {\bf 111}, 157205 (2013).
\bibitem{d46}	D. Stange, S. Wirths, N. von den Driesch, G. Mussler, T. Stoica, Z. Ikonic, J. M. Hartmann, S. Mantl, D. Gr\"{u}tzmacher, and D. Buca, ACS Photonics, {\bf 2}, 1539 (2015).
\bibitem{d47}	K. Zelazna, M. P. Polak, P. Scharoch, J. Serafinczuk, M. Gladysiewicz, J. Misiewicz, J. Dekoster, and R. Kudrawiec, Appl. Phys. Lett., {\bf 106}, 142102 (2015).

\end{thebibliography}
\end{document}